\documentclass[aps, nofootinbib, reprint, preprintnumbers, superscriptaddress, showkeys]{revtex4-2}

\usepackage{amsmath}
\usepackage{amssymb}
\usepackage{tabularx}
\usepackage{graphicx}
\usepackage[usenames,dvipsnames,svgnames]{xcolor}
\usepackage{hyperref}
\usepackage{nameref}
\usepackage[T1]{fontenc}
\usepackage{float}

\hypersetup{dvips,dvipdfm,colorlinks=true,urlcolor=magenta,filecolor=magenta,linktoc=page,citecolor=red,linkcolor=blue,bookmarks=true}

\begin{document}

\title{Evolution of primordial black holes in an adiabatic FLRW universe with gravitational particle creation}

\author{Subhajit Saha}
\email{subhajit1729@gmail.com}
\affiliation{Department of Mathematics, Panihati Mahavidyalaya, Kolkata 700110, West Bengal, India}

\author{Abdulla Al Mamon}
\email{abdulla.physics@gmail.com}
\affiliation{Department of Physics, Vivekananda Satavarshiki Mahavidyalaya (affiliated to the Vidyasagar University), Manikpara-721513, West Bengal, India}

\author{Somnath Saha}
\email{sahasomnath847@gmail.com}
\affiliation{Department of Mathematics, Sree Chaitanya College, Habra 743268, West Bengal, India}


\begin{abstract}

\begin{center}
(Dated: The $22^{\text{nd}}$ February, $2022$)
\end{center}

We study the evolution of primordial black holes (PBHs) in an adiabatic FLRW universe with dissipation due to bulk viscosity which is considered to be in the form of gravitational particle creation. Assuming that the process of evaporation is quite suppressed during the radiation era, we obtain an analytic solution for the evolution of PBH mass by accretion during this era, subject to an initial condition. We also obtain an upper bound on the accretion efficiency $\epsilon$ for $a \sim a_r$, where $a_r$ is the point of transition from the early de Sitter era to the radiation era. Furthermore, we obtain numerical solutions for the mass of a hypothetical PBH with initial mass 100 g assumed to be formed at an epoch when the value of the Hubble parameter was, say, 1 km/s/Mpc. We consider three values of the accretion efficiency, $\epsilon=0.23,0.5$, and $0.89$ for our study. The analysis reveals that the mass of the PBH increases rapidly due to the accretion of radiation in the early stages of its evolution. The accretion continues but its rate decreases gradually with the evolution of the Universe. Finally, Hawking radiation comes into play and the rate of evaporation surpasses the accretion rate so that the PBH mass starts to decrease. As the Universe grows, evaporation becomes the dominant phenomenon, and the mass of the PBH decreases at a faster rate. As argued by Debnath and Paul, the evaporated mass of the PBHs might contribute towards the dark energy budget of the late Universe.

\end{abstract}

\maketitle


\section{Introduction}
\label{sec01}

Primordial black holes (PBHs) refer to hypothetical black holes formed due to fluctuations in the early Universe prior to the big bang nucleosynthesis (BBN) \cite{Novikov0,Hawking1,Chapline1}. Zel'dovich and Novikov \cite{Zel'dovich00} were the first to propose the existence of such black holes in 1966. The evolution of such black holes were later studied by Hawking and Carr \cite{Hawking1,Carr1,Carr2}, and they showed them to be viable candidates for dark matter due to their non-relativistic and effectively collisionless nature. PBHs are classified as massive compact halo objects (MACHOs). PBHs have gained much attention lately \cite{Bird1} after the first detection \cite{Abbot1} of gravitational waves produced by the merger of two $\sim 30~M_{\odot}$ black holes. It has been found from the analysis of posterior data from LIGO and Virgo that the detected mergers of black holes are consistent with the hypothesis of their components being of primordial origin \cite{Sasaki1,Ali1,Clesse1,Clesse2,Garcia-Bellido1,Luca1,Luca2,Wong1}. Moreover, Dolgov et al. \cite{Dolgov1} have found that PBHs with a lognormal mass function fit observational data better as compared to stellar black holes. The study of PBHs is also important because they might provide us with great insights into the initial density fluctuations of the Universe \cite{Carr3,Green1}. PBHs may have been formed just after a mere fraction of a second after the Big Bang because space was inhomogeneous to some extent during that time which might have resulted in certain regions getting hotter and denser as compared to others, and these regions could have collapsed into black holes. Thus, we can conclude that PBHs were formed in the radiation dominated era when the gravitational attraction of denser regions superceeded the radiation pressure \cite{Hawking1,Carr1,Zel'dovich0}. It may, however, be noted that PBHs could also have been formed due to phase transitions in the early universe or due to the collapse of topological defects \cite{Hawking0,Hawking00,Polnarev1,Garriga1,Deng1,Deng2}. The mass of a generated PBH is proportional to the time of its formation and is considered to be a fixed fraction, $\delta$, of the ``standard'' particle horizon mass, $M_{\text{PH}}$ \cite{Carr1,Carr4}
\begin{equation}
M_{\text{PBH}}=\delta M_{\text{PH}}=\frac{4\pi}{3} \delta \rho H^{-3},
\end{equation}
where $\rho$ is the energy density and $H$ is the Hubble parameter. The numerical parameter $\delta$ depends on the details of gravitational collapse. Depending on the epoch of their formation, PBHs could have masses ranging from as low as $10^{-8}$ kgs. to as high as $10^{10}~M_{\odot}$ \cite{Carr1}. Hawking was amazed by the possibility of the formation of such tiny black holes and was motivated to explore their quantum mechanical properties. This led to his discovery that black holes have the ability to evaporate over time by a process which later came to be known as Hawking radiation \cite{Hawking2}. It is quite trivial to note that large black holes would take a longer time to evaporate than small black holes which might have already evaporated or might be doing so at present, depending on their mass. In fact, PBHs evaporate on a time scale $t_{ev}=\frac{5120\pi G^2 M^2}{\hbar c^4}$, which means that PBHs of masses lower than $10^{12}$ kgs. have completely evaporated by now \cite{Carr1}. It is interesting to note that PBHs which have evaporated in early epochs could account for baryogenesis \cite{Barrow0,Majumdar1,Upadhyay1} in the Universe, while PBHs which are long-lived could act as seeds of structure formation or as precursors to supermassive black holes which populate the present epoch \cite{Khlopov1,Dokuchaev1,Khlopov2,Dokuchaev2,Mack1}. The interested readers are referred to some recent reviews which have discussed PBHs extensively \cite{Sasaki2,Carr0,Carr00,Green2,Domingo1}. It is worthwhile to mention here that PBHs, when detected, would provide valuable hints about unknown physics of the very early Universe which is still elusive \cite{Khlopov2,Polnarev2,Belotsky1}. Moreover, as shown by Ketov and Khlopov \cite{Ketov1}, PBHs may also prove useful in probing high-energy scales and supersymmetry theories. A handful of past as well as future observations are expected to constrain the mass and abundance of PBHs, such as Hawking radiation \cite{Hawking000}, lensing of gamma ray bursts \cite{Barnacka1,Katz1}, capture of PBHs by neutron stars \cite{Capela1,Montero-Camacho1}, survival of white dwarfs \cite{Montero-Camacho1,Graham1}, microlensing of stars by the EROS and MACHO surveys, and the Suberu Telescope \cite{Tisserand1,Alcock1,Niikura1,Clesse000} and that of Type Ia supernovae \cite{Miguel1}, CMB anisotropies \cite{Ali-Haimoud00}, and gamma ray signatures from annihilating dark matter detected by the Fermi Gamma-ray Space Telescope \cite{Eroshenko1,Boucenna1}.\\

Accretion of PBHs in the radiation dominated era has raised quite a debate among physicists, with one group suggesting that accretion is not effective in increasing the mass of a PBH sufficiently \cite{Carr1,Zel'dovich0}, while others presenting arguments for contrary possibilities \cite{Majumdar1,Upadhyay1,Hacyan1,Custodio1,Custodio2,Nayak0}. However, in contrast to PBHs in standard Cosmology, enhanced accretion has been shown to be possible for modified gravity theories, such as in braneworld scenario \cite{Guedens1,Majumdar2}, a generalized scalar-tensor model \cite{Majumdar3}, as well as in Brans-Dicke theory \cite{Nayak1} by prolonging the PBH lifetime by significant orders. In the present work, we consider the problem of evolution of PBHs in an FLRW universe driven by particle creation induced by the gravitational field. It is worthwhile to note that the gravitational particle creation mechanism (GPCM) can, in principle, describe the evolutionary history of the Universe in its entirety and is, therefore, often considered to be a viable alternative \cite{Gunzig1,Zimdahl1,Zimdahl2,Lima1,Chakraborty1,Saha1,Mondal1} to both dark energy and modified gravity. Moreover, GPCM has been shown to be thermodynamically stable \cite{Saha2}.\\ 

Schr\"odinger \cite{Schrodinger1} put forward a microscopic theory of the GPCM in an expanding universe. Parker and others \cite{Parker1,Parker2,Birrell1,Mukhanov1} then considered this matter in the context of quantum field theory in curved spacetime. Later, Prigogine et al. \cite{Prigogine1} came up with a macroscopic version. A theory of irreversible thermodynamics in the context of relativistic fluids was pioneered by Eckart \cite{Eckart1}. Landau and Lifshitz \cite{Landau1} introduced a different version of this theory later. It is worth mentioning that this first order theory has several problems such as those related to its stability and causality. A theory of extended irreversible thermodynamics, also sometimes referred to as a second-order theory, is believed to overcome these problems. M\"uller \cite{Muller1} developed a non-relativistic version of this theory. Then, a relativistic description was introduced by Israel \cite{Israel1} and Israel and Stewart \cite{Israel2,Israel3}. The interested reader may see the lectures by Maartens \cite{Maartens1} for a nice review on this topic. Irreversible thermodynamics has also been discussed in the perspective of a covariant theory, initially by Pav\'on et al. \cite{Pavon1} and then by Calvao et al. \cite{Calvao1}.\\ 

Perfect fluids can, in principle, describe most, if not all, of the cosmological models. However, as is well-known, real fluids give rise to dissipation. In fact, several physical processes in Cosmology such as neutrino decoupling, reheating, nucleosynthesis, etc., require a relativistic theory of dissipative fluids. Thus, one needs to employ nonequilibrium thermodynamics in order to have an effective theory of cosmological and astrophysical processes. In a homogeneous and isotropic universe, the only phenomenon that undergoes dissipation is a bulk viscous pressure. This pressure may be created either on account of the coupling of various components of the cosmic substratum \cite{Weinberg1,Straumann1,Schweizer1,Udey1,Zimdahl3} or as a consequence of nonconservation of (quantum) particle number \cite{Zel'dovich1,Zel'dovich2,Murphy1,Hu1}. Our focus will be on the second aspect only.\\ 

For our study, we will consider an adiabatic open system with gravitational creation of particles. Now, following Balfagon's work \cite{Balfagon1}, it can be easily deduced that the thermodynamics of an adiabatic open system with a nonconstant particle number $N$ leads to 
\begin{equation} \label{SN}
dS=\lambda dN,
\end{equation}
where $S$ is the total entropy, while $\lambda = \frac{S}{N}$ is the specific entropy, i.e., the entropy per particle. Eq. (\ref{SN}) is quite simple but, in an adiabatic FLRW universe with a varying particle number, it presents the following two very important results:\\

I. {\bf The second law of thermodynamics (SLT) suppresses the annihilation of particles.} This can be easily verified by dividing Eq. (\ref{SN}) by $dt$, which gives $\dot{S}=\lambda \dot{N}$. As $\lambda >0$, so SLT ($\dot{S} \geq 0$) implies $\dot{N} \geq 0$. This shows that we only need to consider particle creation in an adiabatic FLRW universe where the number of particles changes with time.\\

II. {\bf Specific entropy is an invariant.} Taking the total differential of $S=\lambda N$ and comparing it with Eq. (\ref{SN}), we obtain $d\lambda =0$ which gives $\lambda = \text{constant}$. Note that, if $\lambda$ is a constant for every particle, then the fluid is said to be isentropic. Thus, the isentropy condition is a natural consequence of an adiabatic FLRW universe with a varying number of particles.\\

In this paper, we wish to study the evolution of PBHs which are generally presumed to be formed during the very early Universe. The paper is organized as follows. Section \ref{sec02} outlines the basic equations in an FLRW universe with an isentropic creation of particles induced by the gravitational field. In Section \ref{sec03}, we discuss the evolution of PBHs in such a universe and also analyze some numerical situations. Finally, we present our conclusions in Section \ref{sec04}.


\section{Basic Equations in an FLRW Universe with Isentropic Gravitational Particle Creation}
\label{sec02}

We consider a spatially flat, FLRW universe governed by the line element
\begin{equation}
ds^2=-dt^2+a^2(t)\left[dr^2+r^2(d\theta^2+\mbox{sin}^2\theta d\phi^2)\right],
\end{equation}
where $a(t)$ is the scale factor of the Universe. When bulk viscosity comes into play, the energy-momentum (EM) tensor, $T_{\mu \nu}$, of a relativistic fluid takes the form
\begin{equation}
T_{\mu \nu} = (\rho+p+\Pi)u_{\mu}u_{\nu} + (p+\Pi)g_{\mu \nu},~~~~u_{\mu}u^{\mu}=-1,
\end{equation}
where $\rho$ and $p$ are the energy density and the pressure of the cosmic fluid respectively, $\Pi$ is the bulk viscous pressure, and $u^{\mu}$ is the fluid 4-velocity. We have assumed $c=1$ without any loss of generality. Then, in a flat FLRW universe, the Einstein's field equations $G_{\mu \nu} = 8\pi G T_{\mu \nu}$ give
\begin{eqnarray}
3H^2 &=& 8\pi G \rho, \nonumber \\
\dot{H} &=& -4\pi G(\rho+p+\Pi),
\end{eqnarray}
where $H=\frac{\dot{a}}{a}$ is the Hubble parameter. We further consider that the equilibrium pressure $p$ and the energy density $\rho$ are connected by the equation of state (EoS) defined by $p=(\gamma -1)\rho$, where $\gamma$ is the EoS parameter, taken to be a constant. The EM conservation law ${T^{\mu\nu}}_{;\nu}=0$ gives
\begin{equation} \label{EMCE}
\dot{\rho}+3H(\rho+p+\Pi)=0.
\end{equation}
The fact that we have a varying particle number is accounted for by the equation \cite{Balfagon1}
\begin{equation} \label{PNCE}
\dot{n}+3Hn=n\Gamma,
\end{equation}
where $n=\frac{N}{V}$ is the number density of particles and $\Gamma$ is the rate of creation of particles. Then, using Eqs. (\ref{EMCE}) and (\ref{PNCE}), and noting that $\dot{\rho}=\frac{h}{n}\dot{n}$ \cite{Balfagon1}, where $h$ is the enthalpy density, we establish that
\begin{equation} \label{PI}
\Pi = -\frac{\Gamma}{3H}(\rho +p).
\end{equation}
The above relation is very important since it relates the bulk viscosity and the particle creation rate in a linear fashion under the assumption that the creation of particles is isentropic. Finally, using the Einstein's field equations and Eq. (\ref{PI}), we obtain
\begin{equation} \label{G/3H}
\frac{\Gamma}{3H}=1+\left(\frac{2}{3\gamma}\right)\frac{\dot{H}}{H^2}.
\end{equation}


\section{Evolution of Primordial Black Holes in an FLRW Universe with Isentropic Gravitational Particle Creation}
\label{sec03}

Following the work of Zimdahl \cite{Zimdahl2}, we find that the GPCM can explain the early inflationary era of the Universe if we choose $\Gamma \propto H^2$. This gives
\begin{equation} \label{eq-9}
\Gamma = 3\beta \frac{H^2}{H_r},
\end{equation}
where $3\beta$ is the constant of proportionality and $H_r$ is the Hubble parameter at some fixed scale factor $a_r = a(t_r)$. Plugging Eq. (\ref{eq-9}) into Eq. (\ref{G/3H}) and integrating, we obtain \cite{Zimdahl2,Chakraborty1}
\begin{equation} \label{Hvalue}
H=\frac{H_r}{\beta + (1-\beta)\left(\frac{a}{a_r}\right)^{\frac{3\gamma}{2}}}.
\end{equation}
One can readily observe that as $a \rightarrow 0$, $H \rightarrow \beta^{-1}H_r$. The latter, being a constant, indicates an exponential expansion which must correspond to the inflationary era ($\ddot{a}>0$). Now, if $a \gg a_r$, the second term in the denominator will dominate and we shall have $H \propto a^{-\frac{3\gamma}{2}}$ which represents the matter-dominated era ($\ddot{a}<0$). It is also quite straightforward to determine the relation between the parameters $\beta$ and $\gamma$. In order to achieve that, let us denote $a_r$ as the value of the scale factor at which the acceleration $\ddot{a}$ vanishes. This will mark the transition point from the early de Sitter era onto the radiation era. Since the deceleration parameter $q$ is defined as
$$q=-\frac{\dot{H}}{H^2}-1,$$
we must have $\dot{H}_r=-H_{r}^{2}$ at $a=a_r$. Then, from Eqs. (\ref{G/3H}) and (\ref{eq-9}), we get
\begin{equation}
\beta = 1-\frac{2}{3\gamma}.
\end{equation}
So, for radiation, we have $\gamma=\frac{4}{3}$, which gives $\beta=\frac{1}{2}$. Substituting this value of $\beta$ in Eq. (\ref{Hvalue}), the expression for the Hubble parameter gets reduced to
\begin{equation} \label{Hvalue-1}
H = \frac{2H_r}{1+\left(\frac{a}{a_r}\right)^2}.
\end{equation}

\noindent
Now, for a PBH bombarded by radiation, the rate of change in its mass $M$ caused due to the accretion of radiation is given by \cite{Nayak1}
\begin{equation} \label{Macc}
\dot{M}_{acc.}=4\pi \epsilon R^2 \rho,
\end{equation}
where $R=2GM$ is the radius of the PBH and $\epsilon$ is the accretion efficiency. The value of $\epsilon$ is associated with complex physical processes such as the mean-free paths of the relativistic particles that surround the PBH. Any peculiar velocity of the PBH relative to that of the cosmic frame could enhance the value of $\epsilon$ \cite{Majumdar1,Upadhyay1}. In the literature, it is customary \cite{Guedens1} to consider the rate of accretion to be proportional to the energy density of radiation multiplied by the surface area of the PBH with $\epsilon \sim \mathcal{O}(1)$. This is because the precise value of $\epsilon$ is still not known. On the other hand, the rate of loss of mass of the PBH due to Hawking radiation is given by \cite{Nayak1}
\begin{equation} \label{Meva}
\dot{M}_{eva.}=-4\pi \sigma R^2 T^4,
\end{equation}
where $\sigma = 5.67 \times 10^{-8}~\mbox{W}\mbox{m}^{-2}\mbox{K}^{-4}$ is the Stefan-Boltzmann constant and $T=\frac{1}{8\pi GM}$ is the Hawking temperature. Thus, the rate of change of the PBH mass is given by the sum of the rate of accretion and the rate of evaporation of the PBH,
\begin{equation}
\dot{M}=\dot{M}_{acc.}+\dot{M}_{eva.}.
\end{equation}
Using Eqs. (\ref{Macc}) and (\ref{Meva}), we obtain the rate of change of the PBH mass as
\begin{equation}
\dot{M}=6\epsilon M^2 H^2-\frac{\sigma}{256 \pi^{3} M^2}
\end{equation}
in gravitational units, i.e., in which $G=1$. Noting that $\dot{M}=aH\frac{dM}{da}$, we find that the mass of the PBH follows the evolution equation
\begin{equation} \label{Mevol}
\frac{dM(a)}{da}=\frac{1}{aH(a)}\left[6 \epsilon M(a)^2 H(a)^2 - \frac{\sigma}{256 \pi^3 M(a)^2}\right],
\end{equation}
where $H(a)$ is given by Eq. (\ref{Hvalue-1}). In order to make matters a bit simpler, we now apply a transformation of variables $a \rightarrow a'$ and $H \rightarrow H'$ defined by $a'=a/a_r$ and $H'=H/H_r$ respectively. With these transformations, the evolution equation (\ref{Mevol}) becomes
\begin{equation} \label{Mevol-1}
\frac{dM(a')}{da'}=\frac{\mu}{a'H'(a')}\left[\delta M(a')^2 H'(a')^2 - \eta M(a')^{-2}\right],
\end{equation}
where $\mu=1/H_r$, $\delta=6 \epsilon H_{r}^{2}$, and $\eta=\sigma/256 \pi^3=7.143229048 \times 10^{-12}$. Introducing the expression for $H'=H/H_r$ into Eq. (\ref{Mevol-1}), we obtain
\begin{equation} \label{Mevol-2}
\frac{dM}{da'}=2 \mu \delta \frac{M^2}{a'(1+a'^2)}-\mu \eta \frac{(1+a'^2)M^{-2}}{2a'}.
\end{equation}
This is a highly non-linear ordinary differential equation which we could not solve by hand or even using a sophisticated software such as Maple 13. So, for the time-being, we look for solutions when the process of evaporation is suppressed, or in other words, when the accretion is the dominant phenomenon. This will be achieved during the radiation era so that we can have $a \sim a_r \Leftrightarrow a'=1$ in the second term on the right-hand side of Eq. (\ref{Mevol-2}). This will modify the evaporation term as $\mu \eta/M^2$, which will have a very small positive value even for PBHs of masses as low as $10^{-8}$ kgs. Thus, if we ignore the evaporation term and impose the initial condition
\begin{equation} \label{ic}
M(a_r)=M_r,
\end{equation}
we can have the initial value problem (IVP) given by
\begin{equation} \label{ivp}
\frac{dM}{da'}=2 \mu \delta \frac{M^2}{a'(1+a'^2)},~~~~M(a_r)=M_r
\end{equation}
with $a'=a/a_r$. After separation of the variables, we obtain\footnote{The lower limit of the integral on the right-hand side is $a'=a_r/a_r$ which equals 1.}
\begin{equation} \label{Mevol-3}
\int_{M_r}^{M} \frac{dM}{M^2}=2 \mu \delta \int_{1}^{a'}\frac{da'}{a'(1+a'^2)},
\end{equation}
which on integration gives
\begin{equation} \label{Mevol-4}
M=M_r\left[1-12 \epsilon H_r M_r \left\{\mbox{ln}\left(\frac{\sqrt{2}\left(\frac{a}{a_r}\right)}{\sqrt{1+\left(\frac{a}{a_r}\right)^2}}\right)\right\}\right]^{-1},
\end{equation}
where we have substituted $a'$ by $a/a_r$. Note that this solution is an approximation but it is expected to give pretty accurate values for the PBH mass when $a \sim a_r$. This condition also restricts the accretion efficiency $\epsilon$ as
\begin{equation} \label{epsineq}
\epsilon < \frac{1}{6H_rM_r\mbox{ln}~2}.
\end{equation}

\begin{figure*}[]
\includegraphics[width=0.625\textwidth]{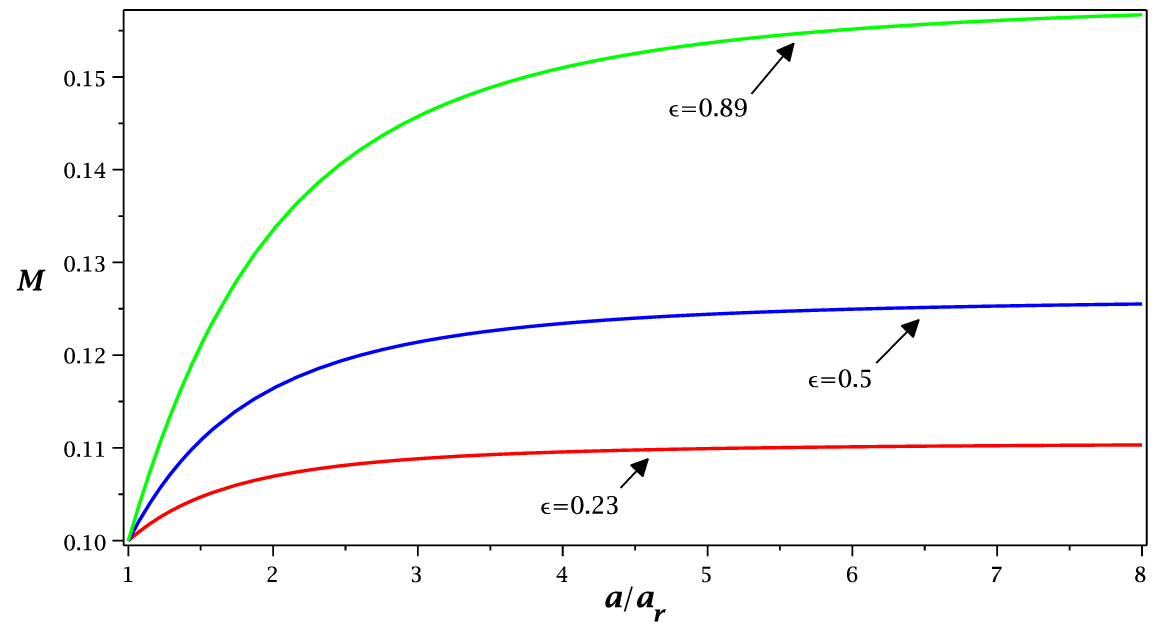}
\caption{Evolution of the PBH mass by accretion against $a/a_r$ with $H_r=1$, $M_r=10^{-1}$, and $\epsilon=0.23, 0.5, 0.89$}
\label{fig}
\end{figure*}

\noindent
It would be nice if we could visualize Eq. (\ref{Mevol-4}) for some numerical choices of the parameters $\epsilon, H_r, M_r$. Let us consider a PBH of initial mass $M_r=100$ g formed at an epoch when the value of the Hubble parameter $H_r$ was, say, 1 km/s/Mpc. We assume three different values of $\epsilon$, $\epsilon=0.23,0.5,$ and $0.89$ in order to have a comparative picture of the effect of accretion efficiency on the evolution of the PBH mass dominated by the process of accretion. We have plotted\footnote{The figure has been produced with Maple 13.} the evolution of the PBH mass by accretion with respect to $a/a_r$ in Figure \ref{fig} for $H_r=1$ and $M_r=10^{-1}$. It is evident from the figure that the mass of the PBH increases a bit faster during the initial stages of its evolution particularly because of the fact that the process of accretion is much more effective during the early radiation phase. It then continues to evolve in a uniform fashion before settling down to some asymptotic value. The effect of $\epsilon$ is quite clearly demostrated by the three curves. As is expected, a higher value of $\epsilon$ leads to a faster evolution of the PBH mass. This type of evolution of the PBH mass during the effective accretion phase is similar to that obtained by other authors such as by Majumdar et al. \cite{Majumdar1} (see Figure 1 on Page 523).\\

Now, we shall look to obtain a numerical solution of Eq. (\ref{Mevol-2}) endowed with the initial condition given by Eq. (\ref{ic}) in order to have a complete picture of evolution of the PBH mass in an FLRW universe with dissipation in the form of isentropic gravitational particle creation. Once again, we consider a PBH of initial mass $M_r=100$ g formed at an epoch when the value of the Hubble parameter $H_r$ was, say, 1 km/s/Mpc. We also assume the same set of values of $\epsilon$ as before, $\epsilon=0.23,0.5,$ and $0.89$. We use the calling sequence
\begin{flushleft}
$dsol:=dsolve(\{ode,ic\}, numeric, output=listprocedure)$ \\
$M:=~eval(M(x), dsol)$
\end{flushleft}
(ode: given ordinary differential equation, ic: given initial condition) available in Maple 13 \cite{Maple13} to obtain a numerical solution. By default, Maple 13 employs the Runge-Kutta Fehlberg numerical method that produces a fifth order accurate solution to an IVP. We tabulate the numerical values of the PBH mass calculated for 14 values of $a/a_r$ in the range $[1,10^4]$ (in other words, $a \in [a_r,10^4 a_r]$) in Table \ref{tab1a}, Table \ref{tab1b}, and Table \ref{tab1c} for $\epsilon=0.23,0.5$, and $0.89$ respectively. All the three scenarios show a similar type of evolution of the PBH mass. The mass of the PBH increases rapidly during the early stages of its evolution due to accretion of radiation being the dominant phenomenon. The accretion process continues till $a/a_r=10^2$ for our PBH, but the rate of accretion gradually slows down. Eventually, the rate of evaporation takes over and gradually becomes the dominant phenomenon so that the mass of the PBH begins to decrease beyond this epoch. A closer look at the numerical values shows that the mass decreases quite rapidly during the later phases of the evolution which indicates a gradual increase in the rate of evaporation as the Universe evolves.

\begin{table}[t] \centering
\caption{Numerical values of the PBH mass $M$ for different values of the normalized scale factor $a/a_r$. We use units in which $c=G=1$ and assume that $H_r=1$, $M_r=10^{-1}$, and $\epsilon=0.23$.\\}
\begin{tabular}{ccc}
\hline
$a/a_r$ 		& & $M$ (in g) \\
\hline \hline
 1   			& & 100.0 \\
 1.25 			& & 102.8 \\
 1.5 			& & 104.7 \\ 
 1.75 			& & 106.0 \\ 
 2  			& & 106.9 \\
 3 				& & 108.8 \\
 5    			& & 109.9 \\
 10   			& & 110.4 \\
$10^2$ 			& & 110.6 \\
$10^3$ 			& & 110.4 \\
$2 \times 10^3$ & & 110.0 \\
$5 \times 10^3$ & & 106.8 \\
$8 \times 10^3$ & & 100.3 \\
$10^4$ 			& & 93.4 \\
\hline \hline
\label{tab1a}
\end{tabular}
\end{table}

\begin{table}[t] \centering
\caption{Numerical values of the PBH mass $M$ for different values of the normalized scale factor $a/a_r$. We use units in which $c=G=1$ and assume that $H_r=1$, $M_r=10^{-1}$, and $\epsilon=0.5$.\\}
\begin{tabular}{ccc}
\hline
$a/a_r$ & & $M$ (in g) \\
\hline \hline
 1   			& & 100.0 \\
 1.25 			& & 106.3 \\
 1.5 			& & 110.8 \\ 
 1.75 			& & 114.0 \\ 
 2  			& & 116.4 \\
 3 				& & 121.4 \\
 5    			& & 124.4 \\
 10   			& & 125.8 \\
$10^2$ 			& & 126.2 \\
$10^3$ 			& & 126.1 \\
$2 \times 10^3$ & & 125.8 \\
$5 \times 10^3$ & & 123.4 \\
$8 \times 10^3$ & & 118.6 \\
$10^4$ 			& & 113.9 \\
\hline \hline
\end{tabular}
\label{tab1b}
\end{table}

\begin{table}[t] \centering
\caption{Numerical values of the PBH mass $M$ for different values of the normalized scale factor $a/a_r$. We use units in which $c=G=1$ and assume that $H_r=1$, $M_r=10^{-1}$, and $\epsilon=0.89$.\\}
\begin{tabular}{ccc}
\hline
$a/a_r$ & & $M$ (in g) \\
\hline \hline
 1   			& & 100.0 \\
 1.25 			& & 111.8 \\
 1.5 			& & 121.0 \\ 
 1.75 			& & 128.1 \\ 
 2  			& & 133.5 \\
 3 				& & 145.7 \\
 5    			& & 153.6 \\
 10   			& & 157.4 \\
$10^2$ 			& & 158.8 \\
$10^3$ 			& & 158.7 \\
$2 \times 10^3$ & & 158.5 \\
$5 \times 10^3$ & & 157.0 \\
$8 \times 10^3$ & & 154.1 \\
$10^4$ 			& & 151.3 \\
\hline \hline
\end{tabular}
\label{tab1c}
\end{table}






\section{Discussion}
\label{sec04}

This paper dealt with a study of the evolution of PBHs in an adiabatic FLRW universe with bulk viscosity which has been assumed to be generated due to a creation of particles induced by the gravitational field. As shown by Zimdahl \cite{Zimdahl2}, the GPCM is able to explain the early inflationary era if the particle creation rate $\Gamma$ is proportional to the square of the Hubble parameter. He determined the Hubble parameter during the radiation era by connecting the two parameters $\beta$ and $\gamma$. The parameter, $\beta$, is a constant of proportionality, while $\gamma$ is the EoS parameter of the barotropic cosmic fluid. We have considered the equations governing the evolution of PBH mass due to accretion by radiation and that due to evaporation by the process of Hawking radiation. The effective evolution of PBH mass is then given by the sum of these two quantities. This equation has been presented as an initial value problem (IVP) and we have been able to obtain an analytic solution for the evolution of the PBH mass through accretion assuming that the evaporation is suppressed during the radiation era. This analytic solution has a logarithmic nature and is an approximation for the PBH mass but is expected to give accurate values for epochs near $a_r$. We have also found an upper bound of the accretion efficiency $\epsilon$ for $a \sim a_r$. For a graphical interpretation, we have considered a hypothetical PBH of initial mass 100 g assumed to have been formed during an epoch when the value of the Hubble parameter was 1 km/s/Mpc. Using Maple 13, we have plotted the mass of the PBH with respect to the normalized scale factor $a/a_r$ for three different choices of $\epsilon$. We observe that the mass of the PBH increases in a rapid fashion during its early evolutionary phase due to a faster accretion rate. The accretion process then continues in a uniform fashion before settling down to an asymptotic value. The effect of the value of $\epsilon$ on the change of mass is quite evident from the figure. Finally, in order to have a complete picture of evolution of the PBH, we have determined numerical values of the PBH mass for 14 values of $a/a_r$ in the range $[1,10^4]$ using Maple 13. These values have been presented in three tables corresponding to the three different values of $\epsilon$. The analysis of these numerical values reveals that the mass of the PBH increases rapidly due to accretion of radiation in the early stages of its evolution. The accretion continues but its rate decreases gradually with the evolution of the Universe. Finally, Hawking radiation comes into play and the rate of evaporation surpasses the rate of accretion so that the PBH mass starts to decrease. As the Universe grows, evaporation becomes the dominant phenomenon and the mass of the PBH decreases at a faster rate. As speculated by Debnath and Paul \cite{Debnath1}, the evaporated mass of the PBHs might contribute towards the dark energy budget of the late Universe. The latest analysis of LIGO/Virgo gravitational wave data by Franciolini and collaborators \cite{Franciolini1} suggests that out of the 47 binary black hole events observed by LIGO/Virgo till date, more than a quarter of these collisions might involve PBHs. However, this fraction depends heavily on the set of assumed astrophysical formation models. The authors in Ref. \cite{Franciolini1} further add that this remarkable possibility involving PBHs could only be verified by minimizing uncertainties in the formation models, and it may finally be confirmed by third generation interferometers. As far as our knowledge is concerned, our present work is the first instance of modelling the evolution of PBHs in a gravitationally induced particle creation scenario. The applications of this model are not apparent immediately because there has been no confirmed candidate of a PBH yet. However, we are hopeful enough to point out the following two possible applications of our model which can be tested by future observations ---\\

(1) An upper bound on the accretion efficiency $\epsilon$ can be obtained from the inequality in Eq. (\ref{epsineq}) once we detect at least one confirmed candidate of a PBH and determine the values of the parameters $H_r$ and $M_r$ from the observations. On the other hand, if $H_r$ is known, then the fact that $\epsilon \sim \mathcal{O} (1)$ will impose bounds on the mass parameter $M_r$. This will give an idea about the initial mass of these PBHs.\\

(2) If the parameters $\epsilon$, $H_r$, and $M_r$ are determined precisely from the observations of a confirmed case of a PBH, then the complete evolution of its mass can be computed using the numerical technique described in this paper.


\begin{acknowledgments}
The author S.S. is grateful to Arindam Kumar Chatterjee and Bibekananda Nayak for introducing him to the excitng field of Primordial Black Holes through some informal discussions and also for drawing his attention to some important references in the field. He would also like to thank Ritabrata Biswas for a discussion on black hole accretion.
\end{acknowledgments}



\end{document}